# Chaotic behavior of ion exchange phenomena in polymer gel electrolytes through irradiated polymeric membrane

Sangeeta Rawat, Barnamala Saha, Awadhesh Prasad and Amita Chandra*

Department of Physics and Astrophysics, University of Delhi, Delhi 110007, India

*Corresponding author Amita Chandra, Department of Physics and Astrophysics, University of Delhi, Delhi 110007, India; E-mail: achandra@physics.du.ac.in; Phone: 91-11-27662295; Fax: 91-11-27667061

**Abstract**

A desktop experiment has been done to show the nonlinearity in the I-V characteristics of an ion conducting electrochemical micro-system. Its chaotic dynamics is being reported for the first time which has been captured by an electronic circuit. Polyvinylidene fluoride-co-hexafluoropropene (PVdF-HFP) gel electrolyte comprising of a combination of plasticizers (ethylene carbonate and propylene carbonate) and salts have been prepared to study the exchange of ions through porous poly ethylene terephthalate (PET) membranes. The nonlinearity of this system is due to the ion exchange of the polymer gel electrolytes (PGEs) through a porous membrane. The different regimes of spiking and non-spiking chaotic motions are being presented. The possible applications are highlighted.

1. Introduction

Porous membranes are thin films or layers in which pores of specified diameters are produced by irradiation of high energy particles and subsequent etching. These membranes are of huge technological importance as they are ideal models of porous systems. Membrane filtration and separation have become common in many branches ranging from microelectronics to biotechnology, pharmaceutical production and water recycling [1-4]. Porous membranes have the intriguing property of ion current fluctuation for a constant voltage across the membrane. It is based on the concept that conducting and non-conducting states represent open and closed configuration of channels, respectively, (same as opening and closing of a gate) [5-6].

Porous solid polymer membranes of PET have excellent mechanical strength and similar to biological channels, they show ion selectivity and inhibition to current flow by protons and divalent cations [7-9]. Many other features of irradiated polymer membranes have been described previously [10-14]. For current flow due to ions, polymer gel electrolytes are an excellent substitute for liquid electrolytes. They comprise of high dielectric constant plasticizer/solvent with different salts (lithium, magnesium, sodium etc.) which are immobilized in a matrix of a host such as PVdF-HFP, poly (methyl methacrylate (PMMA)), polyethylene oxide (PEO) etc. [5].

The present study is on ion migration of polymer gel electrolytes through a porous membrane in the presence of an external potential. Here, PVdF-HFP gel electrolyte has been used which consists of a combination of plasticizers (ethylene carbonate (EC) and propylene carbonate (PC)) and salts (lithium perchlorate and magnesium perchlorate) dissolved in a solvent tetrahydrofuran (THF). While the plasticizers have been added to increase the fluidity of the electrolyte, the solvent is only a medium which facilitates ion motion ($Li^+$ and $ClO_4^-$) and does not contribute to the conductivity value in any significant way [15-20]. The polymer gel electrolytes having different salt content have been separated in a two compartment cell by a porous PET membrane and their current-voltage behaviour has been studied. Further, the current fluctuations due to ion exchange of the polymer gel electrolytes have been investigated.

The main aim of this paper is to investigate the dynamical behavior in ion conducting gel electrolytes through porous PET membrane employing the modified Sprott circuit. The spiking and non-spiking dynamics in a wide range of parameter space has been observed. Within a micro-system, it is difficult to analyze nonlinearity due to the complexities involved in the experimental set-up. Here, a simple set-up is being presented to capture and analyze the non-linearity in such systems via an electronic circuit which can be used to demonstrate the chaotic motion in the laboratory frame-work. Chaotic motion has been studied extensively in the last few decades in a variety of systems, e.g., mechanical, chemical, physical, as well as in social sciences [21-28]. The present study provides a new dimension to the study of electrochemical systems using a simple desk-top experiment.

2. **Experimental**

Polymer membrane (PET) of thickness 12 µm has been irradiated by Kr (250 MeV) ions at a fluence of $4 \times 10^6/cm^2$ to create latent ion tracks in it. Etching conditions of the irradiated foil determine the shape of the pore. To produce nearly cylindrical pores in PET, irradiated foil has been etched in NaOH (1 M) at 40°C. Irradiation of the PET foil followed by etching of the latent tracks breaks some of ester bonds (generating free carboxyl groups) and makes the porous membrane more hydrophilic [27]. The dangling bonds respond to external electric fields and also introduce ion-current rectification mechanism [28].

Two different polymer gel electrolytes (PGE) of 80:20 weight percent (having maximum conductivity) have been prepared to study the ion transport phenomena through the membranes. They are: (i) *Li polymer gel electrolyte*: 1g of PVdF-HFP copolymer has been dissolved in 25 ml THF by stirring at 40°C for 3-4 hours. Then, plasticizers EC: PC (1:1 V/V ratio) have been used and $LiClO_4$ has been mixed to make (1 M) solution of $LiClO_4$. Subsequently, a known amount of this solution has been added to the previous polymer solution and stirred (2 hours) to get the Li polymer gel electrolyte. The prepared lithium polymer gel electrolyte has conductivity 0.03 mS/cm and pH 5 (at 25°C), (ii) *Mg polymer gel electrolyte*: Similarly, PVdF-HFP (1g) copolymer has been dissolved in 25 ml THF by stirring at 40°C for 3-4 hours. Then, 1M solution of $Mg(ClO_4)_2$ has been prepared in EC:PC (1/1 V/V ratio). Its known amount has been dissolved in the polymer solution and stirred for 2 hours to get the Mg polymer gel electrolyte. The prepared magnesium gel electrolyte has conductivity 0.5 µS/cm and pH of 5 (at 25°C). All the chemicals have been purchased from Sigma Aldrich.

3. **Electrical Characterization**

In the present study, two different PGEs having different ionic conductivity and same pH have been filled in the two compartments of the cell separated by a porous PET membrane. Figure 1 gives the schematic diagram of the cell configuration for I-V measurements where, A & B are the compartments with Li and Mg polymer gel electrolyte, respectively; C & D are the platinum electrodes for compartments A and B, respectively. F is the porous PET membrane.

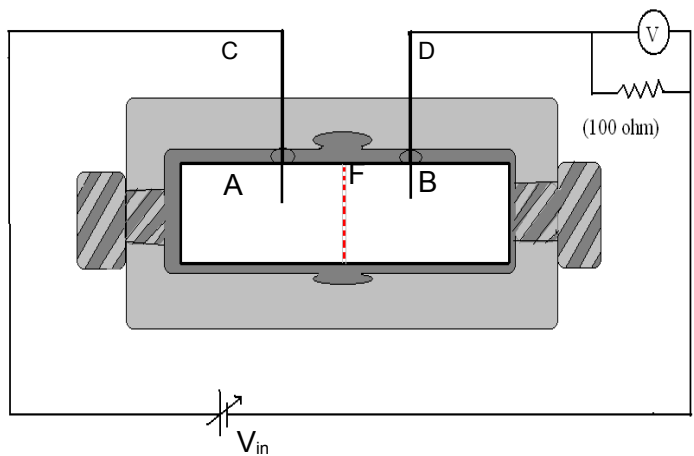

**Figure 1.** Schematic diagram for I-V measurement

## 4. Results

The I-V characteristics have been shown in figure 2 for different configurations: (curve a) Li polymer gel electrolyte on both sides of membrane, (curve b) Mg polymer gel electrolyte on both sides of membrane, (curves c and d) Li PGE on one side and Mg PGE on other side of the cell, having different pore sizes. While in curves a, b and c, the etching time is 3 hours, for curve d, the etching time is 1 hour. Shorter etching time gives smaller pores as compared to the relatively longer etching time. This has been done to show the universality of the observed non-linearity in the Li-Mg system. Note that the non-linearity in the I-V only shifts its position (voltage value) on varying the pore size and therefore, is a property of the system.

The I-V behaviour of this system will not only be affected by the pore size but also by the ion type and conductivity of the polymer gel electrolytes. In case of curves 'a' and 'b', there is no abrupt change in the current as the system contains the same ion type, ion concentration, conductivity and pH values. It also indicates that the magnitude of ion current is higher for Li-Li system as compared to the other two systems because it contains only univalent lithium ions. While in the Mg-Mg system and Li-Mg system, the current magnitude is low because of the lower conductivity of the Mg PGE and the divalent $Mg^{2+}$ ions whose movement is restricted via the PET membrane.

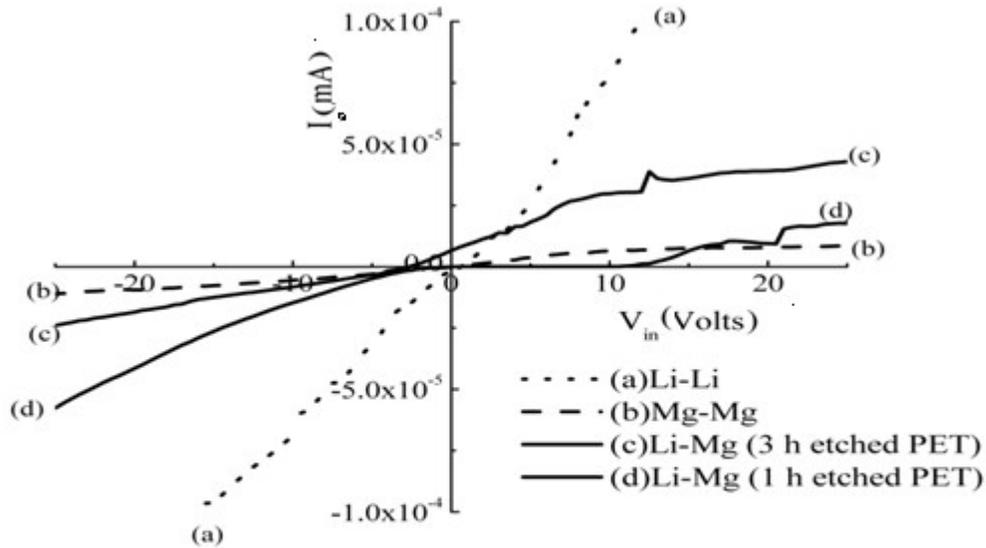

**Figure 2.** I-V characteristics of the polymer gel electrolytes in a two compartment cell separated by a porous PET membrane; (a) Li polymer gel electrolytes on both sides of the cell (dotted line), (b) Mg polymer gel electrolyte on both sides of the cell (dashed line) and (c & d) Li PGE on one side and Mg PGE on the other side (solid line)

For the Li-Mg system separated by a porous PET membrane, due to the biasing, the $Li^+$ ions will preferably flow from the Li PGE compartment to the Mg PGE compartment, hence, giving rise to a current in the forward direction. It must be noted that due to the property of PET membrane described above, $Mg^{2+}$ ions will not cross over to the Li PGE compartment. It is known that the PGEs are predominantly cation conductors with nominal anion mobility [19]. In the present system, due to the favourable biasing, the anions ($ClO_4^-$) in the Mg PGE compartment will start collecting at the porous PET membrane. This will further increase the field for more $Li^+$ flow to the Mg PGE compartment. Due to their large radii and low mobility, only a few anions will be able to cross over to the Li PGE compartment. However, when the applied potential exceeds a critical value (in the present case, ~11V for etching time 3 hour) which is sufficient to pull the anions to the Li PGE compartment, a peak in the conductivity is observed due to the sudden release of charge carriers (anions). For smaller pores size (etching time ~1 hour), this peak is observed at a higher critical voltage (~20 V) as more field it required to pull the anions to the Li PGE compartment when the pores are smaller.

In order to capture the different dynamical behviour due to the nonlinearity in the Li-Mg system, modification has been done in the Sprott's circuit [29], figure 3. The Sprott circuit contains three successive integrators coupled with the nonlinear element G(x) (dotted box). The nonlinear element G(x) (present case: Li-Mg PGE two compartment cell) has been connected in series with the power supply ($V_0$) and diode. An additional component, i.e., the diode has been used because it provides the required bias to the two-compartment cell. When $V_O<V_X$, then the diode is forward biased and the cell does not play any significant role in the circuit. However, when $V_O>V_X$, the diode is reverse biased and the cell is included in the circuit.

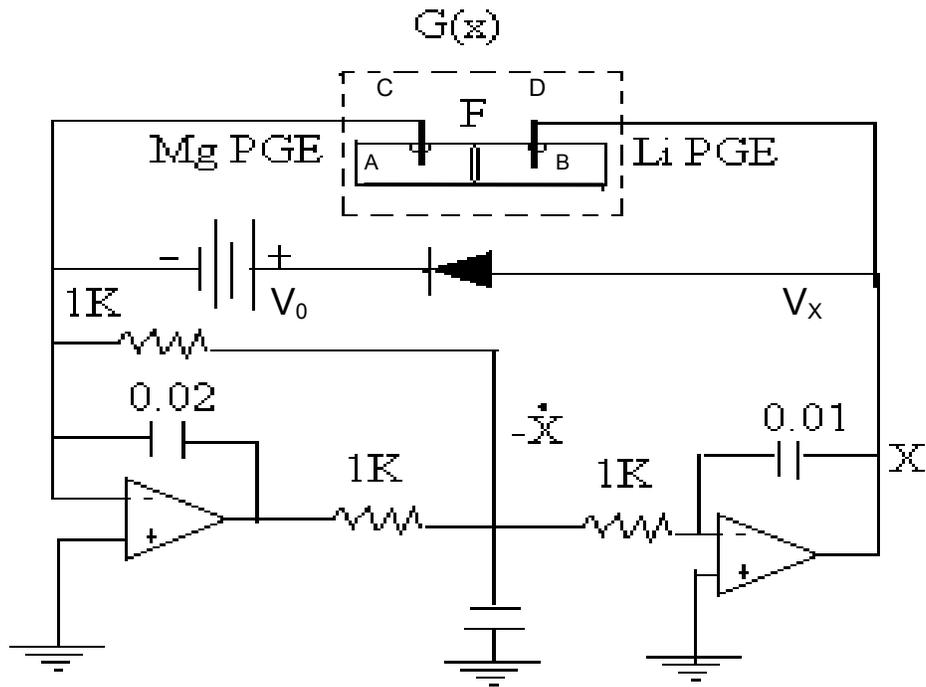

**Figure 3.** Modified Sprott's circuit for trapping the nonlinearity due to ion motion

Using the above configuration, at different external applied potential $V_0$, time series data corresponding to the points X (Volt) and dX/dt have been collected using NI-Lab view and ELVIS bundle. A set of data points have been recorded at uniformly spaced time intervals (3 msec). The bifurcation diagram (using maxima $X_m$ of X, with $V_0$) is shown in figure 4(a). It indicates that there is a sudden drop (around critical applied voltage $V_{oc}$) in the values of $X_m$, as applied voltage ($V_0$) increases in the range of 0-14V. In this voltage range, two sets of time series data, at applied voltages 6V and 10V, have been chosen for further analysis. The bifurcation diagram for a different etching time (~1 hour) is given in figure 4(b) which shows similar behavior as figure 4(a) but for a different $V_{oc}$ value.

In order to see the different dynamics across the transition $V_{oc}$ (figure 2 (a)), the plots of time series of X (volts) with time are shown in figures 5 (a) and (c) before (6V) and after (10V) the transition, respectively. Figures 5(b) and (d) show the corresponding phase space plots (attractors) of 5(a) and (c), respectively. These plots clearly show the occurrence of irregular oscillating motions. Further analysis of these data shows that the motions are chaotic with Lyapunov exponents 0.0076 and 0.035, respectively (given in **table 1**). One of the main difference between these two, before and after the transition $V_{oc}$ is that, in the former regime, motion is non-spiking (Fig. 5(a)) (having typical oscillating behavior) while in the latter regime, motion is spiking (Fig. 5(c)).

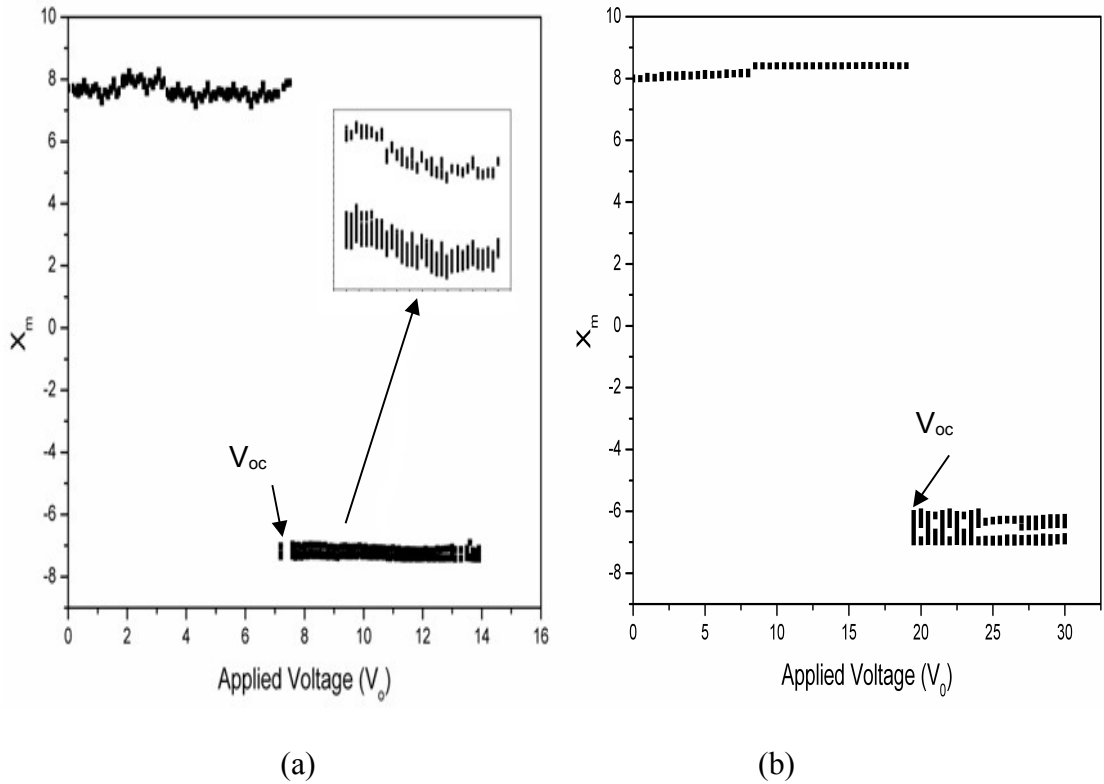

**Figure 4.** Bifurcation diagram with voltage $V_0$ when etched by 1M NaOH for (a) 3hours and (b) 1 hour

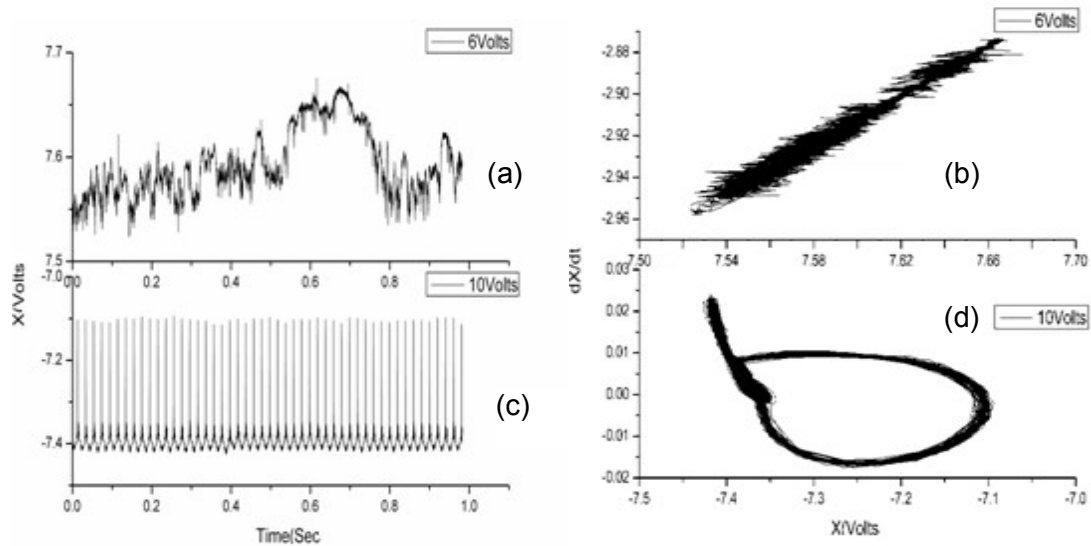

**Figure 5.** Plot of X vs. time for $V_O$ (a) 6V and (c) 10V, and corresponding phase space trajectories (b) and (d), respectively (for 3hrs. etched PET film)

**Table1.** Calculated phase-restructure parameters and Lyapunov exponents of voltage data

| Applied voltage ($V_0$) | Delay (sec) | Embedding dimension | Lyapunov exponent | Error |
|---|---|---|---|---|
| 6volts | 6 | 5 | 0.0076 | $6.4 \times 10^{-4}$ |
| 10 volts | 0.015 | 5 | 0.035 | $1.19 \times 10^{-3}$ |

## 5. Conclusion

The radiation effects on polymers have attracted much attention since the last two decades due to fundamental interest and their wide range application in electronics, medical devices and automobiles. In particular, irradiated and etched PET membranes have been studied extensively due to their ability to mimic biological channels. Our study, using porous PET membranes in an ion exchange system, is summarized below.

From this study, it has been observed that, two polymer gel electrolytes having different conductivities and salt with different ions, when separated by a porous PET membrane (having varying pore sizes), can show sustained ion current fluctuation even if the concentration and pH values are the same. Due to the geometry of the pores in the membrane and the favorable movement of univalent cations as compared to divalent cations through it, the ion current fluctuation leads to a nonlinear I-V characteristic. When the characteristic is further integrated with modified Sprott's circuit, the chaotic behaviour of the present system has been observed. Different dynamical behaviors, *spiking and non-spiking* in current, have also been observed.

The significance of the system reported here is as follows: (i) chaotic behaviour has been observed in an electrochemical system (in-vitro) and (ii) time series analysis of an in-vitro system has been done similar to the in-vivo systems like the biological systems. Some examples of the application of such an ion selective system are: (a) clinical sensor, as the membranes are sensitive to univalent cations, (b) Li ion containing biological fluid analyzer and (c) ion selective electrodes [30-31]. For patients undergoing lithium therapy, accurate and rapid monitoring of the $Li^+$ activity in blood is critical as the gap between its therapeutic and toxic levels is very close. Nowadays, lithium ion selective electrodes are being used in the clinical field. Ion selective electrodes consist of a thin membrane across which only the intended ion can be transported. In the case of a potentiometric electrode, a potential develops in the presence of one ion but not in the presence of a similar concentration of other ions [32-34].

In the present study, polymer gel electrolytes containing two different ions ($Li^+$ and $Mg^{2+}$) with same concentration have been tested with different applied potential and the selectivity of $Li^+$ ion has been confirmed in a broad range of applied potential (-32V to +32V). This indicates the lithium ion selective nature of the polymeric system under study. The aforesaid nature of system coupled with the light weight and flexibility of the porous polymeric membrane makes it a viable material to be used in ion selective devices.

**Acknowledgement**

We wish to thank the UGC, CSIR, DST (Govt. of India) and the University of Delhi for financial support. Our thanks are due to Prof. Ram Ramaswamy for critical comments and Miss Vinita Suyal for contribution in data analysis. The authors are also grateful to Dr. D. Fink and the operators at the Helmholtz Centre for Materials and Energy, Berlin for irradiation of the samples.